\documentclass{article}
\usepackage[dvips]{graphicx}
\usepackage{times,cite}
\usepackage{tabls}

\def\ge{\epsilon}
\def\gg{\gamma}
\def\gd{\delta}

\def\gm{\mu}
\def\gn{\nu}

\def\gp{\pi}

\def\gl{\lambda}
\def\gL{\Lambda}

\def\delmu{\partial_\gm}
\def\delmuu{\partial^\gm}
\def\delnu{\partial_\gn}

\def\delrlmu{\stackrel {\leftrightarrow} {\partial_\gm}}

\def\delrl1{\stackrel {\leftrightarrow} {\partial_1}}

\def\part{\partial}

\def\hlf{\frac{1}{2}}
\def\A0{A^{+}_0}

\def\Psib{\overline{\Psi}}

\def\ulix{\underline{x}}
\def\uliy{\underline{y}}

\newcommand{\nc}{\newcommand}
\nc{\intl}{\int\limits_{-L}^{+L}\!\frac{{\rm d}x^-}{2}}
\nc{\intly}{\int\limits_{-L}^{+L}\!{{{\rm d}y^-}\over\!2}}
\nc{\intlz}{\int\limits_{-L}^{+L}\!{{{\rm d}z^-}\over\!2}}
\nc{\intlu}{\int\limits_{-L}^{+L}\!{{{\rm d}u^-}\over\!2}}
\nc{\intlv}{\int\limits_{-L}^{+L}\!{{{\rm d}v^-}\over\!2}}
\nc{\intv}{\int\limits_{-V}^{}\!{\rm d}^3\ulix}
\nc{\intvy}{\int\limits_{-V}^{}\!{\rm d}^3\uliy}
\nc{\zmint}{\int\limits_{-L}^{+L}\!{{{\rm d}x^-}\over{\!2L}}}
\nc{\zminty}{\int\limits_{-L}^{+L}\!{{{\rm d}y^-}\over{\!2L}}}
\nc{\intp}{\int\limits_{0}^{+\infty}\!{{{\rm d}p^+}\over\!4\gp}}
\nc{\inp}{\int\limits_{0}^{\infty}\!{{{\rm d}p^+}\over{\!2\sqrt{2\gp}}}}
\nc{\inq}{\int\limits_{0}^{\infty}\!{{{\rm d}q^+}\over{\!2\sqrt{2\gp}}}}
\nc{\inpp}{\int\limits_{0}^{\infty}\!{{{\rm d}p^+}\over{\!2p^+\sqrt{2\gp}}}}
\nc{\ink}{\int\limits_{0}^{\infty}\!{\rm d}k^+}
\nc{\inqq}{\int\limits_{0}^{\infty}\!{{{\rm d}q^+}\over{\!2q^+\sqrt{2\gp}}}}
\nc{\insl}{\int\limits_{-L}^{+L}\!{\rm d}x}
\nc{\intex}{\int\limits_{-\infty}^{+\infty}\!{\rm d}x^1}
\nc{\intey}{\int\limits_{-\infty}^{+\infty}\!{\rm d}y^1}
\def\iny2d{\int\limits_{-\infty}^{+\infty}\!{\rm d}^2y}
\def\inz2d{\int\limits_{-\infty}^{+\infty}\!{\rm d}^2z}
\nc{\intep}{\int\limits_{-\infty}^{+\infty}\!{\rm d}p^1}
\nc{\inteq}{\int\limits_{-\infty}^{+\infty}\!{\rm d}q^1}
\nc{\intek}{\int\limits_{-\infty}^{+\infty}\!{\rm d}k^1}
\nc{\intel}{\int\limits_{-\infty}^{+\infty}\!{\rm d}l^1}
\nc{\inty}{\int\limits_{-\infty}^{+\infty}\!{\rm d}y^-}
\nc{\intz}{\int\limits_{-\infty}^{+\infty}\!{\rm d}z^-}
\def\beq{\begin{equation}}
\def\eeq{\end{equation}}
\def\bea{\begin{eqnarray}}
\def\eea{\end{eqnarray}}
\nc{\intx}{\int\limits_{-\infty}^{+\infty}\!\frac{{\rm d}x^-}{2}}
\nc{\intgix}{\int\limits_{-\infty}^{+\infty}\!{{{\rm d}x^-}\over\!2}}
\nc{\intgiy}{\int\limits_{-\infty}^{\infty}\!{{{\rm d}y^-}\over\!2}}

\def\tB{\tilde{B}}
\def\tA{\tilde{A}}
\def\tG{\tilde{G}}

\begin{document}
\title{New operator solution of the Schwinger model in a covariant gauge  
and axial anomaly  
\thanks{Presented at the Light Cone 2012, Krakow, Poland, July 8--13, 2012}
}
\author{L$\!\!$'ubom\'{\i}r Martinovi\u{c}\\ 
BLTP JINR Dubna, Russia \\
and\\ 
Institute of Physics, Slovak Academy of Sciences \\
D\'ubravsk\'a cesta 9, 845 11 Bratislava, Slovakia\\
}
\maketitle
\begin{abstract}
Massless QED(1+1) - the   
Schwinger model - is studied in a covariant gauge. The main new ingredient is   
an operator solution of the Dirac equation expressed directly in terms of  
the fields present in the Lagrangian. This allows us to study in  
detail residual symmetry of the covariant gauge. For  
comparison, we analyze first an analogous solution in the  
Thirring--Wess model and its implication for the axial anomaly arising 
from the necessity to correctly define products of fermion operators via 
point-splitting. In the Schwinger model, one has to define the currents in a 
gauge-invariant (GI) way. Certain problems with their usual derivation  
are identified that obscure the origin 
of the massive gauge boson. We show how to  
define the truly GI interacting currents, reformulate 
the theory in a finite volume and clarify role of the gauge zero mode in 
the axial anomaly and in the Schwinger mechanism. A trasformation  
to the Coulomb gauge representation is suggested 
along with  
ideas about how to correctly obtain other properties of the model.   
\end{abstract}
\section{Introduction}
The Schwinger model \cite{Schw} 
is a prototype gauge model, studied in 
hundreds of papers using all kinds of techniques. A natural question concerns  
therefore the necessity to perform another study of this subject. What can be 
added/improved in our understanding of the physics of the model? 
Surprisingly enough, no generally accepted picture of the physical content of 
the model is available and some controversies persist. This is nicely 
illustrated by comparing two representatives of the vast literature on the 
subject: the seminal work by Lowenstein and Swieca \cite{LS} and its  
mathematically rigorous reexamination \cite{Str}.  
Both start from the operator solution in Landau gauge in terms of "building  
block" fields, namely using Ansaetze for $A^\mu(x)$, $J^\mu(x)$ and 
$J^\mu_5(x)$ that depend on fields not present in the original 
Lagrangian. The second paper disagrees with the former one in such issues as 
the choice of a minimal set of dynamical variables (the  
correct "intrinsic field algebra" and existence of "bleached states") and 
also with the conclusions about the vacuum structure of the model.   

In this contribution, we will make an attempt to clarify the 
situation using a Hamiltonian approach that reformulates dynamics 
consistently in terms of true degrees of freedom, namely the free fields. 
In particular, we will focus  
on a few overlooked aspects related to truly gauge-invariant (GI) definitions 
of the interacting currents and the consequent issues of the axial anomaly and 
dynamical generation of the boson mass. We start 
with a brief discussion of the related Thirring-Wess (TW) model for comparison.     
The key element is the explicit solution of the Dirac equation in the 
covariant gauge in terms of the fields present 
in the starting Lagrangian, i.e. without using auxiliary 
fields that obscure some aspects of the problem.  
Interacting currents can then be calculated directly from the 
known solutions in a regularized form ("point-splitting") in both models.   
The difference is that one has to insert an exponential of the line integral 
of the gauge field to compensate violation of the local symmetry in the gauge   
model. The corresponding divergences of the currents should therefore 
differ in the two models. However, this is not the case in the usual 
treatment! The explanation will be given and the key ideas and elements of  
the full solution of the model will be formulated. 

\section{The Thirring-Wess model}
The model \cite{TW,Brown} is defined by the classical Lagrangian 
\beq
{\cal L}=\frac{i}{2}\Psib\gg^\mu\delrlmu \Psi-\frac{1}{4}\tG_{\mu\nu}
\tG^{\mu\nu} 
+\mu_0^2\tB_\mu \tB^\mu -eJ_\mu \tB^\mu,~~~\tG_{\mu\nu}=\delmu \tB_\nu-
\partial_\nu \tB_\mu.  
\label{TL}
\eeq
The original solutions were either based on indirect methods using Ansaetze in 
terms of auxiliary fields or certain redundant definitions of 
"gauge-invariant" operators. No reliable solution of the model seems to  
have been obtained so far.  

The above Lagrangian leads to the set of coupled field equations (Dirac+Proca)  
\beq   
i\gg^\mu\partial_\mu\Psi(x)= e\gg^\mu \tB_\mu(x)\Psi(x),  
~~~
\delmu \tG^{\mu\nu} +\mu_0^2 \tB^\nu = eJ^\nu.   
\label{dipro}
\eeq  
Taking $\partial_\nu$ of the Proca eq. yields $\delmu \tB^\mu = 0$. With this 
condition, the Dirac eq. is solved in terms of $\tB^0(x)$ and 
the free massless fermion field $\psi(x),  
\gg^\mu\delmu \psi = 0$:   
\beq  
\Psi(x)=\exp\Big\{-\frac{ie}{2}\gg^5\intey \ge(x^1-y^1)\tB^0(y^1,t)\Big\}
\psi(x).  
\label{TWsol} 
\eeq   
Here $\ge(x)=\theta(x)-\theta(-x)$. 
Normal-ordering of the exponential is understood.  
With the notation 
$\hat{k}.x \equiv E(k^1)t-k^1x^1,~E(k^1)=\sqrt{k^2_1 + \mu_0^2},
~E(p^1) = \vert p^1\vert $,     
the quantum field expansions of the independent field 
variables of the model are  
\bea    
&&B^0(x) = \int_{-\infty}^{+\infty}\frac{dk^1}
{\sqrt{4\gp E(k^1)}}
\big[a(k^1)e^{-i\hat{k}.x} + a^\dagger(k^1)e^{i\hat{k}.x}\big], \\  
&&\psi(x) = \frac{1}{\sqrt{2\pi}}\intep \big\{b(p^1)u(p^1)
e^{-i\hat{p}.x} + 
d^\dagger(p^1)v(p^1)e^{i\hat{p}.x}\big\}. 
\eea
The quantization rules are 
\beq
\big[a(p^1),a^\dagger(q^1)\big]=   
\{b(p^1),b^\dagger(q^1)\} = \{d(p^1),d^\dagger(q^1)\} = 
\gd(p^1-q^1). 
\label{qrul}
\eeq
The Fock vacuum is defined as  
$a(k^1) \vert 0 \rangle = b(k^1)\vert 0 \rangle = d(k^1)\vert 0 \rangle = 0.$  
The massless spinors are   
$u^\dagger(p^1) = \big(\theta(-p^1),\theta(p^1)\big),~
v^\dagger(p^1) = \big(-\theta(-p^1),\theta(p^1)\big)$.
The component $B^1(x)$ is determined from the operator relation 
$\delmu B^\mu=0$. 

The product of two fermion operators is regularized by the  
point-splitting: 
\beq   
J^\mu(x) = \Psi^\dagger(x+\frac{\ge}{2})\gg^0\gg^\mu \Psi(x-
\frac{\ge}{2}),~~~  
J^\mu_5(x) = \Psi^\dagger(x+\frac{\ge}{2})\gg^0\gg^\mu \gg^5\Psi(x-
\frac{\ge}{2}).    
\label{Bcure}
\eeq    
Using    
$\psi^\dagger(x+\frac{\ge}{2})\gg^0 \gg^\mu (\gg^5)\psi(x-\frac{\ge}{2}) 
= :\psi(x)^\dagger \gg^0 \gg^\mu (\gg^5)\psi(x): 
- \frac{i}{2\gp} 
Tr\big(\frac{\gg^\alpha \ge_\alpha \gg^\mu (\gg^5)}{\ge^2}\big)$ 
as well as the symmetric 
limit $s~lim _{\ge \rightarrow 0}~\frac{\ge^\mu \ge^\nu}{\ge^2} = 
1/2 g^{\mu\nu}$, we find:  
\bea  
J^\mu(x) = j^\mu(x) + \frac{e}{\gp}\tilde{B}^\mu(x),~~~ 
J_5^\mu(x) = j_5^\mu(x) + \frac{e}{\gp}\ge^{\mu\nu}\tilde{B}_\nu(x).  
\label{Bcur}
\eea  
$j^\mu(x)$ and $j_5^\mu(x)$ are the (normal-ordered) free currents. 
The expression in the exponential contains a term of 
order $O(\ge)$ which cancels the singularity in the free-field contraction.  
In this way, a finite quantum correction is generated. 
The vector current is obviously conserved, while the axial "anomaly" $a(x)$ is 
equal to 
\beq 
\delmu J^\mu(x) = a(x) \equiv \frac{g}{2\gp}\ge^{\mu\nu}\tilde{G}_{\mu\nu}(x).
\label{axano}
\eeq 
It is remarkable that this is precisely the result known from 
the Schwinger model  
although no exponential of the integral over gauge field was inserted ! 

The Proca equations  
become, due to the relation $\delmu \tB^\mu = 0$ and the form of the 
interacting current, also soluble.  
Defining the retarded Green's function by $\big(\delmu\delmuu + \mu^2 \big)
D_R(x-y) = \gd^{(2)}(x-y),~\big(\delmu\delmuu + \mu^2\big)B^\mu(x) = 0$, 
where $\mu^2 = \mu_0^2 - \frac{e^2}{\gp}$, the resultant equation  
$\delmu\partial^\mu \tB^\nu(x) + \mu^2 B^\nu(x) = j^\nu(x)$ 
can indeed be inverted as:   
\beq
\tB^\nu(x) = B^\nu(x) + e\iny2d D_R(x-y)j^\nu(y). 
\eeq
Then the  Hamiltonian can be expressed in terms of the above  
independent fields. 
In the final-volume treatment, also the zero  
mode $b^1(t)$ will play a role. 
The questions  
to be studied are diagonalization of the Hamiltonian deriving thereby the true 
physical vacuum state of the model and a potential chiral symmetry breaking.

\section{Schwinger model in the Landau gauge}  
We will start from the classical Lagrangian  
\bea 
&&{\cal L}=\frac{i}{2}\Psib\gg^\mu\delrlmu \Psi-\frac{1}{4}F_{\mu\nu}F^{\mu\nu} 
-eJ_\mu A^\mu -G(x)\delmu A^\mu + \hlf(1-\gg)G^2(x), \nonumber \\ 
&&~~~~~~F_{\mu\nu}=\delmu A_\nu -\delnu A_\mu,~ J^\mu(x) = \Psib(x)
\gg^\mu\Psi(x), 
\label{SML}
\eea  
that contains two additional terms with respect to the usual QED$(1+1)$. 
For arbitrary $\gg$, these terms restrict the theory to an arbitrary Lorentz 
(covariant) gauge (replacing the usual  
term $-\frac{\gl}{2} \big(\delmu A^\mu(x)\big)^2$)  
in which neither the condition $\delmu A^\mu(x)=0$ nor the Maxwell equations  
can be satisfied at the operator level:  
\bea 
\delmu F^{\mu\nu}(x) = eJ^\nu(x) - \partial^\nu G(x),~~ 
\delmu A^\mu(x) = (1-\gg)G(x).
\label{Maxwalt}
\eea 
The auxiliary field $G(x)$ satisfies $\delmu\partial^\mu G(x) = 0$. 
Choosing $\gg=1$, the gauge condition is 
satisfied at the operator level and the solution of the Dirac equation  
$i\gg^\mu\partial_\mu\Psi(x)= e\gg^\mu A_\mu(x)\Psi(x)$ 
is completely analogous to the TW model case:  
\beq
\Psi(x)=\exp\Big\{-\frac{ie}{2}\gg^5\intey \ge(x^1-y^1)A^0(y^1,t)\Big\}\psi(x), 
~~~\gg^\mu \delmu \psi = 0.
\label{SMsol} 
\eeq
In order to guarantee that we are working with the original theory, 
the condition on physical states $G^{(+)}(x)\vert phys\rangle = 0$,   
has to be used, It generalizes the usual Gupta-Bleuler condition 
$\delmu A^{(+)\mu}\vert phys \rangle = 0$.   
Again, the vector and axial-vector currents have to be calculated via the 
point-splitting. 
It is important to keep in mind that the gauge freedom has been restricted 
only partially, the  
Lagrangian is still invariant with respect to gauge transformations 
parametrized by the gauge function obeying 
\beq
\delmu\partial^\mu \gL(x) = 0  \Rightarrow \partial_0^2 \gL = \partial_1^2 
\gL~~\Rightarrow \frac{\partial_0}{\partial_1}\gL = \frac{\partial_1}
{\partial_0}\gL.  
\label{lorgf} 
\eeq

The conclusion about appearance of a massive vector boson in the theory with 
gauge invariance crucially depends on the axial anomaly. For its derivation,  
one starts from the  "gauge-invariant" definition of the axial current (see 
\cite{PS, Strlec}, e.g.), i.e. one inserts the gauge-field exponential 
to the point-split product of the fields:   
\beq   
J^\mu_{(5)}(x) = \Psi^\dagger(x+\frac{\ge}{2})\gg^0\gg^\mu (\gg^5) 
\exp{\big\{-ie\int\limits
_{x-\ge/2}^{x+\ge/2}dz_\mu A^\mu(z)\big\}}\Psi(x-
\frac{\ge}{2}).   
\label{Bcuree}
\eeq   
No gauge fixing has been done in (\ref{Bcuree}). Both currents are formally 
GI under 
\beq
\Psi(x) \rightarrow e^{ie\gL(x)}\Psi(x),~~~A^\mu(x) \rightarrow A^\mu(x) - 
\partial^\mu \gL(x).  
\label{ginv}
\eeq
The vector current takes the form  
\beq   
J^\mu(x) = \Big[:\Psi^\dagger(x)\gg^0\gg^\mu\Psi(x): +  
\underbrace{\Psi^\dagger(x+\frac{\ge}{2})\gg^0\gg^\mu \Psi(x-\frac{\ge}{2})}
\Big]\Big[1-ie\ge_\nu A^\nu(x)\Big].
\label{calve}
\eeq 
Note that in this derivation, the fermion and gauge fields are taken as 
independent and the free-field contraction has been used. 
The result is precisely
\beq
J^\mu(x) = j^\mu(x) + \frac{e}{\gp}A^\mu(x),~~ 
J^\mu_5(x) = j^\mu_5(x) + \frac{e}{\gp}\ge^{\mu\nu}A_\nu(x), 
\label{vcu}
\eeq
i.e. gauge-NON-invariant expressions! 
This fact is hidden since one usualy calculates directly 
the divergence which gives the "familiar" (gauge-invariant) anomaly 
(\ref{axano}). 
How should one understand the above contradiction? To answer this question, 
let us calculate the anomaly carefully using our Landau-gauge operator 
solution (\ref{SMsol}).  
We have to take into account that the general transformation law $A^\mu 
\rightarrow A^\mu - \partial^\mu\gL$ becomes     
$A^0(x) \rightarrow A^0(x) - \partial_0 \gL(x),~~~
\delmu\partial^\mu \gL=0$ in our gauge and   
{\bf this completely determines the transformation law for the interacting 
fermion field} since the free fermion field $\psi(x)$ does not transform: 
\bea 
\Psi(x) \rightarrow \exp\big\{\frac{ie}{2}\gg^5\intey \ge(x^1-y^1)\partial_0 
\gL(y^1,t)\big\}\Psi(x) \equiv \exp\big\{\frac{ie}{2}\gg^5 \frac{\partial_0}
{\partial_1}\gL\big\}\Psi(x). 
\label{gisol} 
\eea 
The point of course is that 
$\Psi(x)$ and $A^\mu(x)$ are not independent. We have to modify the  
"gauge exponential" in such a way that the (split) currents  
are invariant under the  
specific transformations (\ref{gisol}). 
The correct form of the current is  
\beq    
J^\mu_{(5)}(x) = \Psi^\dagger(x+\frac{\ge}{2})\gg^0\gg^\mu (\gg^5)
\exp{\big\{-ie
\gg^5\ge_{\mu\nu}A^\mu(x)\ge^\nu\big\}}\Psi(x-
\frac{\ge}{2}),   
\label{truegi}
\eeq
since the gauge variations in the exponential cancel. 
The interacting currents found in this way  
coincide with the free currents!
The implication is no anomaly and therefore no Schwinger mechanism! 
This really looks like a very strange result. 

To understand the situation better, let us analyze the residual gauge symmetry 
and interacting currents in an infared-regularized framework by  
restricting $-L \leq x^1 \leq L$ and imposing (anti)periodic boundary 
conditions for the free fields:
\bea   
\!\!\!\psi(t,-L) = -\psi(t,L),~A^\mu(t,-L) = A^\mu(t,L)  
\Rightarrow A^\mu(x) = 
A^\mu_N(x) + A^\mu_0(t).
\label{bc}
\eea   
$A^\mu_0(t)$ is the gauge field zero mode (ZM). 
The gauge transformations have the form    
\bea 
&&A^\mu_N(x) \rightarrow A^\mu_N(x) - \partial^\mu \gL_N(x), \nonumber \\
&&A^0_0(t) 
\rightarrow A^0_0(t)-\partial_0 \gL_0(t),~~~A^1_0(t) \rightarrow A^1_0(t) + 
\partial_1 \gL_0(t) = A^1_0(t).
\label{fvgt}
\eea 
The gauge conditions are $\partial_0 A^0_N(x) + \partial_1 A^1_N(x) = 0$,   
$A^0_0(t)=0$. 
The Dirac eq. and its solution is  
\bea 
&&\!\!\!\!\!\!\!\!\!i\gg^0\partial_0\Psi + i\gg^1\partial_1\Psi =
e\big(\gg^0 A^0_N 
-\gg^1A^1_N
\big)\Psi 
-e\gg^1A^1_0(t)\Psi, 
\label{fvdir}  
\eea
\beq
\Psi(x)=\exp\Big\{ie\gg^5\Big[\int\limits_{t_0}^t d\tau 
A^1_0(\tau) 
-\int\limits_{-L}^{+L} dy^1 
\ge_N(x^1-y^1)A^0_N(x^1-y^1)\Big]\Big\}\psi(x). 
\label{fvsol}
\eeq 
The GI currents have the form 
\beq
J^\mu_{(5)}(x) = \exp\big\{-ie\gg^5\ge^0A^1_0(t)\big\}\psi(x+\frac{\ge}{2})
\gg^0\gg^\mu(\gg^5)\psi(x-\frac{\ge}{2}).   
\label{corrcur}
\eeq
Contraction in the discrete basis has the same singular structure as in the 
continuum and we obtain  
$J^\mu(x) = j^\mu(x) + \frac{e}{\gp}(0,A^1_0(t)),~J^\mu_5(x) = j^\mu_5(x) + 
\frac{e}{\gp}(A^1_0(t),0)$. 
Both currents are gauge invariant since $A^1_0(t)$ component is GI by itself. 
Then  
\bea 
&&\delmu J^\mu(x) = \delmu j^\mu(x) + \frac{e}{\gp} 
(0,\partial_x A^1_0(t))=0, 
\\
&&\delmu J^\mu_5(x) = \delmu j^\mu_5(x) + 
\frac{e}{\gp}(\partial_0 A^1_0(t),0) = \frac{e}{\gp}\partial_0 
A^1_0(t) \ne 
0. 
\label{corrdiv}
\eea
From the ZM part of the Maxwell eq. one directly has  
\beq
\partial_0^2 A^1_0(t) = -\frac{e^2}{\gp} A^1_0(t).   
\label{zmschw}
\eeq
We have thus found that the Schwinger mechanism works only in the zero-mode 
sector, where it gives rise to the    
massive Schwinger boson with the mass $\mu^2=\frac{e^2}{\gp}$.  

Next steps in the analysis will involve  
an introduction of the indefinite-metric space, explicit solution of the  
Maxwell equations and a derivation of the Hamiltonian in terms of  
independent field variables along with a study of its invariances (chiral 
symmetry, large gauge transformations). 
For example, the (modified) Maxwell equations 
$\delmu\delmuu \tA^\nu = e j^\nu - \partial^\nu G$  
will be inverted as 
\beq
\tA^\mu(x) = A^\mu(x) + e\iny2d D_{0R}(x-y)j^\mu(y) - \iny2d D_{0R}(x-y)
\delmuu G(y). 
\label{maxol}
\eeq
Presence of the unphysical fields $A^\mu(x)$ in (\ref{maxol}) is related  
to the residual gauge freedom, which can be removed   
by means of a unitary transformation to the Coulomb gauge representation  
\cite{KH,Lenz}. 
It is also necessary to find a mechanism for the vacuum degeneracy in the 
present approach. Here the gauge zero mode and its residual (large) 
gauge symmetry may play a role (note that the covariant gauge admits  
transformations with the gauge function of the form $cx^1$). 
These topics are presently under study.  
\begin{flushleft}
{\bf Acknowledgements:} This work was supported by the grant 
VEGA 2/0070/2009. 
\end{flushleft}

\end{document}